\documentclass[a4paper,11pt]{article}
\usepackage{t1enc}
\usepackage[utf8]{inputenc}
\usepackage[english]{babel}
\usepackage{graphicx}
\usepackage{dsfont}
 \usepackage{amsfonts, amsmath, amssymb, amscd, amsthm, bbm}

\title{\bf Why ``noncommutative common causes'' \\don't explain anything\\[1.5ex] \normalsize \rm A comment on \\G\'abor Hofer-Szab\'o's \textit{Noncommutative causality in algebraic quantum field theory} }

\author{Dustin Lazarovici \\[1.1ex] Ludwig-Maximilians-University Munich, Department of Mathematics \\[1.2ex] \small dustin.lazarovici@math.lmu.de}\date{ }

\begin{document}
\maketitle

\begin{abstract}
\noindent In my commentary, I will argue that the conclusions drawn in the paper \textit{Noncommutative causality in algebraic quantum field theory}\footnote{The paper, alongside this commentary, will appear in \textit{New Directions in the Philosophy of Science} (proceedings of the PSE workshops held in 2012), edited by Maria Carla Galavotti, Dennis Dieks, Wenceslao J. Gonzalez, Stephan Hartmann, Thomas Uebel, Marcel Weber, Springer (2014). The same results have been published, in greater detail, by Hofer-Szab\'o and collaborators in a series of other papers, for instance Hofer-Szab\'o and Vecserny\'es (2013a, 2013b).}
by G\'abor Hofer-Szab\'o are incorrect. As proven by J.S. Bell, a local common causal explanation of correlations violating the Bell inequality is impossible. 

\vspace{0.2in}

\end{abstract}

\begin{quotation} \textit{``Let me guess. He pulled a lost in translation on you?''}\\

\vspace{-0.1in} \hspace{1.8in} \footnotesize{-- from the movie \textit{Ocean's Twelve} (2004)}\\\end{quotation}

What is the meaning of Bell's theorem? What are its implications for which it was dubbed, and rightfully so, ``the most profound discovery of science'' (Stapp 1975)? In brief, Bell's theorem tells us that certain statistical correlations between space-like separated events that are predicted by quantum mechanics and observed in experiment imply that \textit{our world is non-local}. More precisely, it tells us that those correlations are \textit{not locally explainable}, meaning that they cannot be accounted for by any local candidate theory since the frequencies predicted by a local account would have to satisfy a certain inequality -- the Bell, respectively the CHSH inequality -- that is empirically violated in the pertinent scenarios. Any candidate theory that correctly predicts the violation of the Bell inequality must therefore describe some sort of direct influence between the correlating events, even if they are so far apart that they cannot be connected by a signal moving at maximum the speed of light. Hence we say that the principle of \textit{locality} or \textit{local causality} is violated in nature. 

The genius of Bell's argument lies in its simplicity and in its generality. Bell's theorem is not about quantum mechanics or quantum field theory or \textit{any} theory in particular, it is not confined to the ``classical'' domain or the quantum domain or a relativistic or non-relativistic domain, it is a \textit{meta-theoretical} claim, excluding (almost) all possibilities of a local explanation for the statistical correlations observed in EPR-type experiments.\footnote{The only additional assumption entering the derivation of the Bell inequalities is the  so-called \textit{no-conspiracy} assumption, meaning that certain parameter choices involved in the experiments are assumed to be ``free'' or random, that is, not predetermined in precisely such a way as to arrange for apparently non-local correlations. See (Bell 1990) for details.} 
Admittedly, a statement about nature can never reach the same degree of rigor as a theorem of pure mathematics for there is always an issue of connecting formal concepts to ``real-world'' concepts. Bell, however, was one of the clearest thinkers of the 20th century and his analysis, unobscured by the misunderstandings of some of his later commentators, is perfectly precise and conclusive in this respect.\footnote{A beautiful presentation of his analysis can be found in (Bell 1981) and (Bell 1990), the original version of the theorem is (Bell 1964). For a more recent discussion, see (Goldstein et. al 2011) or (Maudlin 2011). The most common misunderstandings are addressed, for instance, in (Norsen 2006) or (Goldstein et.al. 2011).} It is against this background that contributions to the subject have to be evaluated.\\

The paper \textit{Noncommutative causality in algebraic quantum field theory} by G\'abor Hofer-Szab\'o, that I was gratefully given the opportunity to comment on, seems to be an offspring of a research project started about one and a half decades ago by Miklos R\'edei (R\'edei 1997) and concerned with the question whether in Algebraic Quantum Field Theory\footnote{Algebraic Quantum Field Theory is often referred to as \textit{Local Quantum Field theory}, which is a quite unfortunate double-use of terminology. ``Locality'' in quantum field theory usually refers to the postulate of ``microcausality'' or ``local commutativity'', requiring that operators localized in space-like separated regions of space-time commute. This, however, is very different from the concept of Bell-locality as discussed above. In AQFT, local commutativity assures the impossibility of faster-than-light signaling, the theory nevertheless contains \textit{non-local} correlations between space-like separated events due to the non-local nature of the quantum state which is defined as a functional on the entire ``net'' of operator algebras, all over space-time.} correlations between space-like separated events (in particular such violating the Bell inequality) have local explanations in terms of ``common causes''. You see, what worries me about this research program is that it seems to suggest that the status of Bell's theorem is not yet clear, that the issue of non-locality is not yet settled, because somehow the technical details of Algebraic Quantum Field Theory could turn out to matter, and type III von-Neumann algebras could turn out to matter, and noncommutativity could turn out to matter. But that would be incorrect; none of this really matters.\\

Contrary to what I've just so emphatically stated, Mr. Hofer-Szab\'o makes quite an astonishing announcement. He claims that by committing ourselves to the framework of AQFT and by ``embracing noncommuting common causes'' we can achieve what Bell's theorem would seem to exclude, namely to provide a ``local (joint common causal) explanation for a set of correlations violating the Bell inequalities''. Although such a statement will certainly make a huge impression on people who believe that noncommutativity holds the one great mystery of quantum physics, we should pause for an instant to assess its plausibility. 

Physical events, or ``causes'' and ``effects'', whatever we might mean by that, are certainly not the kind of thing that can either commute or not commute. Operators, I grant, can commute or not commute, and so can perhaps elements of lattices with respect to certain set-theoretic operations. Bell's theorem, however, doesn't care about any of this. Bell's argument is solely concerned with the predictions that a candidate theory makes for the probabilities (relative frequencies) of certain physical events, not with the mathematical structure that it posits to make those predictions or represent those events. So how could it be possible to avoid the consequences of Bell's theorem by denying ``commutativity'', which hasn't been among its premises in the first place?\\

Let me try to explain what I think the result presented in the paper of Hofer-Szab\'o actually consists in and why I think it's completely missing the point as concerning the issue of local causality. Contrary to what is being suggested in the paper, the existence of a ``commuting/noncommuting (weak/strong) (joint) common cause system'' according to its definitions 2 and 3 is \textit{not sufficient} for a local (common causal) explanation of correlations between space-like separated events. Such an explanation would at least be required to \textit{reproduce} the statistical correlations that it was set out to explain. The kind of ``explanation'' that the author provides \textit{doesn't do this}. 

As his paper correctly states, the statistics for the events $A_i, B_i$ are different whether the state is first projected on the possible ``common causes'' (since that's what happens when we compute $\phi \circ E_c$) or not. Most notably, the probabilities for the correlated events after the ``occurrence'' (more correctly: measurement) of  ``noncommuting common causes'' (the right-hand-side of eq. \eqref{noteq} below) satisfy the Bell inequality -- in accordance with Bell's theorem -- whereas the statistical correlations that the author \textit{claims to explain} violate Bell's inequality. Note that in the case where $A_i, B_j$ don't commute with $C_k$ we will generally find that
\begin{equation}\label{noteq} \phi(AB) \neq \sum_{k} \frac{\phi(C_k AB C_k)}{\phi(C_k)} \phi(C_k). \end{equation}

\noindent This is, I assume, a familiar fact (if you find the notation confusing, write $\langle \psi | C_k A_i C_k | \psi \rangle$ instead of $\phi(C_k A_i C_k)$, and so on). In particular, there is nothing deep or mysterious or metaphysically interesting about it, if only we appreciate the fact that the right-hand-side of \eqref{noteq} does not describe the same physical situation in which the system remains undisturbed in the common past of $A$ and $B$, but that the projection on the common cause system (indeed one could think of a measurement of an observable $C$ with spectral decomposition $\lbrace C_k \rbrace$) affects (decoheres) the quantum state in a way that can influence subsequent \mbox{(measurement-)}events. In our case, it will simply destroy the EPR-correlations, so that violations if the Bell inequality don't occur at all. In particular, the correlations described by the left-hand-side of \eqref{noteq} are \textit{not explained} by the right-hand-side of \eqref{noteq} since the two probability distributions are different.

In (Hofer-Szab\'o and Vecserny\'es 2012), the authors explicitly acknowledge this point, yet respond by saying that ``\textit{the definition of the common cause does \emph{not} contain the requirement (which our classically informed intuition would dictate) that the conditional probabilities, when added up, should give back the unconditional probabilities [...] or, in other words, that the probability of the correlating events should be built up from a finer description of the situation provided by the common cause.}'' (p.20)\footnote{I have to emphasize that the corresponding statement is slightly different -- and slightly more plausible -- in the published version (Hofer-Szab\'o and Vecserny\'es 2013a, p. 414), possibly as a result of our exchange on the matter. I leave here the quote from the preprint, which was available to me by the time my commentary was written, and which I still believe to reflect a fundamental misconception on part of Hofer-Szab\'o and collaborators.} \\

Although this doesn't strike me as a particularly strong argument, it may be a good starting point for adding a few remarks and highlighting some of the disagreements between the Hofer-Szab\'o and myself. 

\begin{enumerate}
\item As I see it, the problem here is not with probability theory (``that the conditional probabilities, when added up, should give back the unconditional probabilities''), but rather with the assumption that the ``common cause system'' provides a ``finer description'' of the \textit{same} physical situation. The fact that $C + C^\perp = \mathds{1}$ in terms of operators does not imply that it makes no \textit{physical} difference whether \textit{any} of the ``events'' occur, or \textit{none}. There is a difference between a physical situation in which a photon can either pass or not pass a polarization filter and a physical situation with no polarization filter at all.  
\item Even if we accepted the premise of the answer, it would not resolve the issue. The interesting question concerning correlations between space-like separated events is whether they can be explained by some sort of local  ``mechanism'' (I'm using ``mechanism'' broadly here). Bell's theorem states that this is impossible if the Bell inequality is violated. The fact that Hofer-Szab\'o presents us a local mechanism that produces \textit{different} correlations that do \textit{not} violate Bell's inequality seems quite irrelevant in this context. 

\item Despite the tone of the paper suggesting a certain naturalness or inevitability to the concepts it explores, we should keep in mind that it was the author himself who has chosen to \textit{redefine} the ``common cause principle''  for the needs and purposes of AQFT (or rather, who has chosen to follow R\'edei 1997 while admitting noncommuting operators). Hence, when confronted with the objection that the very concept he defined is unsubstantial because it lacks a certain crucial property, he can hardly defend himself by pointing out that the concept lacks this property by definition. 

As far as I was able to understand from this and other publications (e.g. R\'edei 1997, Hofer-Szab\'o and Vecserny\'es 2013a, 2013b), the whole reasoning behind the concept of a ``common cause in AQFT'', on which the author's work is crucially based, is that the common cause principle formulated by Hans Reichenbach is somehow ``classical'' and that there is a canonical way to translate or generalize it to the framework of algebraic quantum field theory. However, leaving aside whether Reichenbach's common cause principle is at all the relevant concept in this context (since his discussion had a different focus), to characterize it as ``classical'' strikes me as a rather confused and confusing statement. The common cause principle is a meta-physical concept, formulated in terms of (what some people call) ``classical probability theory''. However, the word ``classical'' in ``classical probability theory'' shouldn't be confused with the same adjective in the term ``classical (i.e. Newtonian) mechanics''. It doesn't refer to a particular \textit{physical theory} that can be empirically tested and falsified, but to a well-founded mathematical framework expressing a certain \textit{way of reasoning} about nature. It is possible, of course, to borrow Reichenbach's definition and replace the probabilistic events, assumed to be modeled on a classical probability space, by projections in local algebras and the (classical) probability measure by a ``quantum state'',  yielding a value between $0$ and $1$ when evaluated on such projections; but there's no reason to believe that this procedure must yield a meaningful notion of ``common causes'' or ``common cause explanations'' in the context of \textit{any} theory. 
Of course, there are people who believe that quantum theory is and has always been about replacing ``classical probability spaces'' by so-called ``quantum probability spaces''. Nevertheless, I would think that, for the sake of a meaningful and enlightening philosophical discussion, we will have to do better. 
In any case, I would insist that, if we work under this general hypothesis, the fact that certain results we obtain may strike us as counterintuitive or even logically inconsistent need not necessarily reflect some sort of quantum weirdness in nature; it may just as well reflect a lack of imagination or understanding on our side to appreciate that quantum physics is \textit{not} always about putting little hats on capital A's and B's and C's to turn them into operators.\footnote{About common controversies or misconceptions regarding the relevance of ``quantum logic'', ``quantum probabilities'' or ``noncommutativity'' for the issue of non-locality, see also (Goldstein et.al. 2011).}

\end{enumerate}
\noindent In my opinion, much of the confusion in the paper stems from its commitment to a particular jargon that insists on using familiar and intuitive terms (``events'', ``common causes'', etc.) with a non-standard meaning (usually referring to operators). Thus, I find it very helpful to drop this jargon altogether and discuss the situation in the framework of good old-fashioned quantum mechanics to see what the result obtained by Hofer-Szab\'o actually consists in.

\noindent Let's consider the usual spin-singlet state

\begin{equation*}\frac{1}{\sqrt{2}}\bigl(\lvert \uparrow \rangle_1 \otimes \lvert\downarrow \rangle_2 - \lvert\downarrow \rangle_1\otimes \lvert \uparrow \rangle_2\bigr),\end{equation*} 
giving rise to EPR-correlations between the results of spin-measurements on two entangled particles, which can be performed simultaneously by Alice and Bob. In the sense promoted in the paper, a ``noncommuting joint common causal explanation'' would, for instance, consist in the following: after the particles leave the EPR-source, we perform a z-spin measurement on particle 1, taking place in the common past of the measurements of Alice and Bob, who are free to choose between certain orientations other than the z-direction. This (chronologically) first measurement now provides a (non-trivial) ``noncommuting joint common cause system'' in the sense of Hofer-Szab\'o, namely
\begin{equation*}\Bigl\lbrace C= \lvert \uparrow \rangle\langle \uparrow\rvert \otimes \mathds{1}, \; C^\perp = \lvert \downarrow \rangle\langle \downarrow\rvert \otimes \mathds{1}\Bigr\rbrace.\end{equation*}

\noindent Obviously, the probabilities for the outcomes of the succeeding measurements will now split (which, the author insist, is the defining property of a ``common cause''). Just as obviously, the first measurement would simply destroy (decohere) the singlet-state so that the outcomes of the spin-measurements by Alice and Bob will no longer be correlated in a way that violates the Bell- or CHSH-inequality.

The result presented in the paper \textit{Noncommutative causality in algebraic quantum field theory}, although more general and technically more sophisticated, doesn't do anything more than this. I leave it to the reader to judge its explanatory value.

\section*{References} 
\begin{list}
{ }{\setlength{\itemindent}{-15pt}
\setlength{\leftmargin}{15pt}}

\item{Bell, J.S. (1964), ``On the Einstein-Podolsky-Rosen paradox'', reprinted in Bell, J.S. (2004) \textit{Speakable and Unspeakable in Quantum Mechanics}. Cambridge: Cambridge Univ. Press, pp. 14--21.}

\item{Bell, J.S. (1981), ``Bertlmann's socks and the nature of reality'', reprinted in Bell, J.S. (2004) \textit{Speakable and Unspeakable in Quantum Mechanics}. Cambridge: Cambridge Univ. Press, pp. 139--158.}

\item{Bell, J.S. (1990), ``La nouvelle cuisine'', reprinted in Bell, J.S. (2004) \textit{Speakable and Unspeakable in Quantum Mechanics}. Cambridge: Cambridge Univ. Press, pp. 233--248.}

\item{Stapp, H.P. (1975), ``Bell's theorem and world process'', \textit{Il Nuovo Cimento} 40B(29):270-276.}   

\item{Goldstein, S. et.al. (2011) ``Bell's theorem'', \textit{Scholarpedia, 6(10):8378}, \\http://www.scholarpedia.org/article/Bell's\_theorem.}

\item{Norsen, T. (2006) ``Bell Locality and the Nonlocal Character of Nature'', \textit{Found. Phys. Letters}, 19(7): 633-655.} 

\item{Maudlin, T. (2011) \textit{Quantum Non-Locality and Relativity}, Third Ed., Malden, Oxford: Wiley-Blackwell.}

\item{Hofer-Szab\'o, G. and Vecserny\'es, P. (2012) ''Bell Inequality and common causal explanation in algebraic quantum field theory.'' \textit{Preprint: http://philsci-archive.pitt.edu/9101}}.

\item{Hofer-Szab\'o, G. and Vecserny\'es, P. (2013a) ``Bell inequality and common causal explanation in algebraic quantum field theory,'' \textit{Studies in the History and Philosophy of Modern Physics} 44 (4), 404--416.}

\item{Hofer-Szab\'o, G. and Vecserny\'es, P. (2013b) ``Noncommutative Common Cause Principles in algebraic quantum field theory'', \textit{Journal of Mathematical Physics} 54.}

\item{R\'edei, M. (1997) ''Reichenbach's Common Cause Principle and Quantum Field Theory,'' \textit{Found. Phys.} 27: 1309--1321.}

\end{list}

\end{document}